\documentclass[12pt]{article}
\usepackage{epsfig}
\usepackage{graphicx}  % standard LaTeX graphics tool
\newcommand{\tg}{\mathop{\rm tg}\nolimits}

\def\choosen{\atopwithdelims..}

\begin{document}

\begin{center}
{\bf\Large On charmonia and charmed mesons photoproduction\\
 at high energy}

\vspace{3mm}
{\sl V.A.Saleev}\footnote{Email: saleev@ssu.samara.ru} \\
{Samara State University, Samara, 443011, Russia and \\
Samara Municipal Nayanova University, Samara, 443001, Russia}\\ and\\
{\sl D.V.Vasin}\footnote{Email: vasin@ssu.samara.ru}\\
{Samara State University, Samara, 443011, Russia}
%
%\authorrunning{V.A.Saleev and D.V.Vasin}
%

%\maketitle              % typesets the title of the contribution

\end{center}

\begin{abstract}
We compare the predictions of the collinear parton model and the
$k_T$-factorization approach in $J/\Psi $ and $D^\star$ meson
photoproduction at HERA energies. It is shown that obtained in the
both approaches $D^\star$ meson spectra over $p_T$ and $\eta $ as
well as $J/\Psi$ meson  $p_T$- and $z$-spectra are very different.
The predictions obtained in the $k_T$-factorization approach are
agree with the experimental data well. We also predict the
nontrivial $p_T$-dependence of the the spin parameter
$\alpha(p_T)$ in the $J/\Psi$ photoproduction.
\end{abstract}

\section{Hard processes in the parton model and $k_T$-factorization approach}

Nowadays, there are two approaches which are used in a study of
the charmonia and charmed mesons photoproduction at high energies.
In the conventional collinear parton model \cite{1} it is
suggested that hadronic cross section, for example, $\sigma
(\gamma p \to c\bar cX,s)$, and the relevant partonic cross
section $\hat \sigma (\gamma g\to c\bar c,\hat s)$ are connected
as follows
\begin{equation}
\sigma ^{PM}(\gamma p\to c\bar cX,s)=\int {dx}
G(x,\mu ^2)\hat \sigma (\gamma g\to c\bar c,\hat s)\mbox{,}
\end{equation}
where $\hat s=xs$, $G(x,\mu ^2)$ is the collinear gluon
distribution function in a proton, $x$ is the fraction of a proton
momentum, $\mu ^2$ is the typical scale of a hard process. The
$\mu ^2$ evolution of the gluon distribution $G(x,\mu ^2)$ is
described by DGLAP evolution equation \cite{2}. In the so-called
$k_T$-factorization approach hadronic and partonic cross sections
are related by the following condition \cite{3,4,5}:
\begin{equation}
\sigma ^{KT}(\gamma p\to c X,s)=\int {\frac {dx}{x}} \int {d\vec
k_T^2}\int {\frac{d\varphi}{2\pi}}\Phi (x,\vec k_T^2,\mu ^2)\hat
\sigma (\gamma g^\star \to c\bar c,\hat s,\vec k_T^2)\mbox{,}
\end{equation}
where $\hat \sigma (\gamma g^\star \to c\bar c,\hat s,\vec k_T^2)$
is the $c\bar c$-pair photoproduction cross section on off
mass-shell gluon, $k^2={k_T}^2=-\vec k_T^2$ is the gluon
virtuality, $\hat s=xs-\vec k_T^2$, $\varphi $ is the azimuthal
angle in the transverse XOY plane between vector $\vec k_T$ and
the fixed OX axis. The unintegrated gluon distribution function
$\Phi (x,\vec k_T^2,\mu ^2)$ can be related to the conventional
gluon distribution by
\begin{equation}
xG(x,\mu ^2)=\int\limits_{0}^{\mu ^2} {\Phi (x,\vec k_T^2,\mu
^2)d\vec k_T^2}\mbox{,}
\end{equation}
where $\Phi (x,\vec k_T^2,\mu ^2)$ satisfies the BFKL evolution
equation \cite{6}. In formulae~(2) the four-vector of a gluon
momentum is presented as follows:
\begin{equation}
k=x p_N+k_T\mbox{,}
\end{equation}
where $k_T=(0,\vec k_T,0)$, $p_N=(E_N,0,0,E_N)$ is the four-vector
of a proton momentum. At the $x\ll 1$ the off mass-shell gluon has
dominant longitudinal polarization along the proton momentum.
Taking into account the gauge invariance of a total amplitude
involving virtual gluon we can write the polarization four-vector
in two different forms:
\begin{equation}
\varepsilon ^\mu (k)=\frac{k_T^\mu}{|\vec k_T|}
\end{equation}
or
\begin{equation}
\varepsilon ^\mu (k)=-\frac{x p_N^\mu}{|\vec k_T|}\mbox{.}
\end{equation}
As it will be shown above formulae (5) and (6)
give the equal answers in calculating of squared amplitudes under
consideration.

Our calculation in the parton model is down using the GRV LO
\cite{7} parameterization for the collinear gluon distribution
function $G(x,\mu^2)$. In the case of the $k_T$-factorization
approach we use the following parameterizations for an
unintegrated gluon distribution function $\Phi(x,\vec
k_T^2,\mu^2)$: JB by Bluemlein \cite{8}; JS by Jung and Salam
\cite{9}; KMR by Kimber, Martin and Ryskin \cite{10}. The detail
analysis of the evolution equations lied in a basis of the
different parameterizations is over our consideration. To compare
different parameterizations we have plotted their as a function of
$x$ at the fixed $\vec k_T^2$ and $\mu^2$ in Fig.1 and as a
function of $\vec k_T^2$ at the fixed $x$ and $\mu^2$ in Fig.2.

Note, that all parameterizations of an unintegrated gluon
distribution function describe the data from HERA collider for the
structure function $F_2(x,Q^2)$ well \cite{8,9,10}.

\section{$D^\star$ meson photoproduction in LO QCD.}

The photoproduction of the $D^\star$ meson was studding
experimentally by H1 and ZEUS Collaborations at HERA ep-collider
($E_e=27.5$ GeV, $E_N=820$ GeV) \cite{11,12}. Because of the large
mass of a $c$-quark usually it is assumed that $D^\star$ meson
production may be described in the fragmentation approach
\footnote{The another approach based on recombination scenario was
suggested recently in \cite{14}} \cite{13}, where
%
%\begin{equation}
%\frac{d\sigma }{dp_T}(\gamma p\to D^\star X)=\int {D_{c\to D^\star }}(z,\mu ^2)\frac{dz}{z}
%\cdot \frac{d\sigma }{dp_T'}(\gamma p\to cX,p_T'=\frac{p_T}{z})
%\end{equation}
%
\begin{equation}
\sigma(\gamma p\to D^\star X,p)=\int {D_{c\to D^\star }}(z,\mu ^2)
\sigma(\gamma p\to cX,p_1=p/z)dz
\end{equation}
and $D_{c\to D^\star }(z,\mu ^2)$ is the universal fragmentation
function of a $c$-quark into the $D^\star$ meson at the scale $\mu
^2=m_D^2+p_T^2$. The fraction of the $D^\star$ produced by a
$c$-quark as measured by OPAL Collaboration
 \cite{15},
$$
\omega _{c\to D^\star }=\int\limits_{0}^{1} {D_{c\to D^\star }(z,\mu ^2)}dz=0.222\pm 0.014\mbox{,}
$$
has been used in our LO QCD calculations to normalize the
fragmentation function.

The Peterson \cite{13} fragmentation function was used as a
phenomenological factor:
\begin{equation}
D_{c\to D^\star }(z,\mu _0^2)=N \frac{z (1-z)^2}{[(1-z)^2+\epsilon z]^2}\mbox{.}
\end{equation}
In the high energy limit or in the case of a massless quark one
has following relation for the four-vectors $p=z p_1$, however in
the discussed here process the $D^\star$ meson energy is not so
large in compare to $M_{D^\star}$ and the following prescription
was used
\begin{equation}
\vec p=z \vec p_1
\end{equation}
together with the mass-shell condition for the $c$-quark energy
and momentum $E_1^2=\vec p_1^2+m_c^2$. We have used
$\epsilon=0.06$ as a middle value between two recent fits of
$D^\star$ meson spectra in $e^+e^-$-annihilation, which based on
massive charm ($\epsilon=0.036$)\cite{16} and massless charm
($\epsilon=0.116$)\cite{17} calculations. The squared matrix
element for the subprocess $\gamma g^\star\to c\bar c$ after
summation over a gluon polarization accordingly (6) may be written
as follows \cite{4,5,18}:
\begin{equation}
\overline{|M|^2}=16 \pi ^2 e_c^2 \alpha _s \alpha \cdot (\hat
s+\vec k_T^2)^2 \biggl[ \frac{\alpha _1^2+\alpha _2^2}{(\hat
t-m_c^2)(\hat u-m_c^2)} -\frac{2 m_c^2}{\vec k_T^2}\left(
\frac{\alpha _1}{\hat u-m_c^2}-\frac{\alpha _2}{\hat
t-m_c^2}\right) ^2\biggr]
\end{equation}
where $\hat s$, $\hat t$ and $\hat u$ are usual Mandelstam
variables,
$$
\alpha _1=\frac{m_c^2+\vec p_{1T}^2}{m_c^2-\hat t}\mbox{, }\alpha
_2=\frac{m_c^2+\vec p_{2T}^2}{m_c^2-\hat u}\mbox{,}
$$
$\vec p_{1T}$ and $\vec p_{2T}$ are the transverse momenta of $c$-
and $\bar c$-quarks, $\vec k_T=\vec p_{1T}+\vec p_{2T}$.

Using formulas (5) for a BFKL gluon polarization four-vector we
can rewrite (10) in the another form:
\begin{eqnarray}
\overline{|M|^2}&=&\frac{16 \pi ^2 e_c^2 \alpha _{s} \alpha }{
     (m_c^2 - \hat t)^2(m_c^2 - \hat u)^2}
     \biggl[ m_{c}^2 \biggl( -2m_c^6 - 4m_c^2\vec p_{1T}^2\vec k_T^2 + m_c^2\vec k_T^4
     + 8\vec p_{1T}^2\vec k_T^4 +\nonumber\\&&+
     3\vec k_T^6 +
     (4m_c^4 + 12\vec p_{1T}^2\vec k_T^2 + 5\vec k_T^4)\hat s -
     (3m_c^2 - 4\vec p_{1T}^2 - 3\vec k_T^2)\hat s^2 + \hat s^3 \biggr)
     +\nonumber\\&&+
     \biggl( 8m_c^6 + 8m_c^2\vec p_{1T}^2\vec k_T^2 - 2m_c^2\vec k_T^4
     - 4\vec p_{1T}^2\vec k_T^4 - \vec k_T^6 -
     12m_c^4\hat s - 4\vec p_{1T}^2\vec k_T^2\hat s -\nonumber\\&&- \vec k_T^4\hat s + 6m_c^2\hat s^2
      - \vec k_T^2\hat s^2 -
     \hat s^3\biggr) \hat t - \biggl( 4\vec p_{1T}^2\vec k_T^2 -\vec k_T^4 + 3(-2m_c^2
     +\hat s)^2\biggr) \hat t^2 +\nonumber\\&&+
     4\biggl( 2m_c^2 - \hat s\biggr) \hat t^3 - 2\hat t^4
     -
     4|\vec p_{1T}| \biggl( |\vec k_T| \cos (\varphi) \Bigl( -2m_c^6
     -\nonumber\\&&- \Bigl( \vec k_T^2 - \hat s - 2\hat t \Bigr) \hat t\Bigl( 2 m_c^2-\hat u\Bigr)
     +
     m_c^4\Bigl( \vec k_T^2 + 3\hat s + 6\hat t\Bigr) +
     m_c^2\Big( 3\vec k_T^4 + \hat s^2 +\nonumber\\&&+ \vec k_T^2(4\hat s - 2\hat t) - 6\hat s\hat t
      - 6\hat t^2\Bigr)\Bigr)
     + |\vec p_{1T}| \cos (2\varphi )\Bigl( m_c^4\vec k_T^2 + \vec k_T^2\hat t\Bigl( 2 m_c^2 - \hat u\Bigr)
     -\nonumber\\&&-
     m_c^2\Bigl( 2\vec k_T^4 + \hat s^2 + \vec k_T^2(3\hat s + 2\hat
     t)\Bigr)\Bigr)
     \biggr) \biggr] \mbox{,}
\end{eqnarray}
where $\varphi$ is the angle between $\vec p_{1T}$ and $\vec k_T$.

In the last case (11) it is easy to find the parton model limit:
\begin{equation}
\lim\limits_{|\vec k_T|\to 0} \int\limits_{0}^{2 \pi} {\frac {d\varphi}{2 \pi}}
\overline{|M|^2}=\overline{|M_{PM}|^2}\mbox{,}
\end{equation}
where
$$\vec p_{1T}^2=\vec p_{2T}^2=\frac {(\hat u-m_c^2)(\hat t-m_c^2)}{\hat s}-m_c^2 \mbox{,}$$
and
%
%This is famous formulae, but we can not use it.
%\begin{eqnarray}
%\overline{|M_{PM}|^2}=16 \pi ^2 e_c^2 \alpha _{s} \alpha \cdot
%\biggl[ -4\biggl( \frac{m_c^2}{m_c^2-\hat
%t}+\frac{m_c^2}{m_c^2-\hat u}\biggr) ^2 +4\biggl(
%\frac{m_c^2}{m_c^2-\hat t}+\frac{m_c^2}{m_c^2-\hat u}\biggr)
%\\+\frac{m_c^2-\hat u}{m_c^2-\hat t}+\frac{m_c^2-\hat t}{m_c^2-\hat u}
%\biggr]
%\end{eqnarray}
%
\begin{eqnarray}
\overline{|M_{PM}|^2}&=&-\frac{16 \pi ^2 e_c^2
\alpha_{s}\alpha}{((m_c^2 - \hat t)^2 (-m_c^2 + \hat s + \hat
t)^2)} \biggl[-2 m_c^8 + 8 m_c^6 \bigl(\hat s + \hat t\bigr)
-\nonumber\\&&- \hat t \bigl(\hat s^3 + 3 \hat s^2 \hat t + 4\hat
s\hat t^2 + 2\hat t^3\bigr) + m_c^2\bigl(\hat s^3 + 6 \hat s^2\hat
t + 8 \hat t^3 + 4 \hat s\hat t (3 \hat t + \hat u)\bigr)
-\nonumber\\&&- m_c^4 \bigl(7 \hat s^2 + 12 \hat t^2 + 4 \hat s (4
\hat t + \hat u)\bigr)\biggr]
\end{eqnarray}

\section{$D^\star$ meson photoproduction at HERA}

In  this part we will compare our results obtained with leading
order matrix elements for the partonic subprocess $\gamma g\to
c\bar c$ in the conventional parton model as well as in the
$k_T$-factorization approach with data from HERA ep-collider. The
data under consideration taken by the ZEUS Collaboration
\cite{11}. Inclusive photoproduction of the $D^{\star\pm}$ mesons
has been measured for the photon-proton center-of-mass energies in
the range $130 < W < 280$ GeV and the photon virtuality $Q^2 < 1$
GeV$^2$. At low $Q^2$ the cross section for $ep\to eD^\star X$ are
related to $\gamma p$ cross section using the equivalent photon
approximation \cite{19}:
$$
d\sigma_{ep}=\int{\sigma_{\gamma p}}\cdot f_{\gamma
/e}(y)dy\mbox{,}
$$
where $f_{\gamma /e}(y)$ denotes the photon flux integrated over
$Q^2$ from the kinematic limit of $Q^2_{min}=m_e^2 y^2/(1-y)$ to
the upper limit $Q^2_{max} = 1$ GeV$^2$, $y=W^2/s$, $s=4 E_N E_e$,
$E_N$ and $E_e$ are the proton and electron energies in the
laboratory frame.

The exact formulas for $f_{\gamma /e}(y)$ is taken from \cite{20}:
$$
f_{\gamma /e}(y)=\frac{\alpha}{2
\pi}\biggl[\frac{1+(1-y)^2}{y}\log{\frac{Q^2_{max}}{Q^2_{min}}}+2
m_e^2 y
\bigl(\frac{1}{Q^2_{min}}-\frac{1}{Q^2_{max}}\bigr)\biggr]\mbox{.}
$$
The limits of integration over $y$ are
$y_{max\choosen{min}}=W^2_{max\choosen{min}}/s$. In our
calculations we used formulas for differential cross section in
the following form:
\begin{eqnarray}
\frac{d\sigma(ep \to eD^\star X)}{d\eta dp_{T}}&=&\int dy
f_{\gamma /e}(y)\int{\frac{dz}{z}D_{c\to
D^\star}(z,\mu^2)}\int{\frac{d\varphi}{2 \pi } } \int{d\vec k_T^2}
\frac{\Phi(x,\vec k_T^2,\mu^2)}{x} \times\nonumber\\&&
\times\frac{2 |\vec p_1| |\vec p_{1T}|}{E_1 (W^2-2 E_N
(E_1-p_{1z}))}\cdot\frac{|\bar M|^2}{16 \pi x W^2}
\end{eqnarray}
The differential cross section as a function of the $D^\star$
pseudorapidity, which is defined as $\eta=-\ln(\tg
\frac{\theta}{2})$, where the polar angle $\theta$ is measured
with respect to the proton beam direction, is shown in Fig.3 where
the kinematic ranges for the $D^\star$ meson transverse momentum
are $2 < p_T < 12$ GeV, $4 < p_T < 12$ GeV, $6 < p_T < 12$ GeV,
correspondingly from up to down.

We see that the results of calculations performed in the collinear
parton with LO in $\alpha_s$ matrix element need additional
$K$-factor $(K\approx 2)$ to describe the data. The value of this
$K$-factor is usual for a heavy quark production cross section in
the relevant energy range. Opposite the results obtained in
$k_T$-factorization approach are agree with the data well.
Especially in case of the JB parameterization of a unintegrated
gluon distribution function. The difference between the
predictions obtained with the various parameterizations is about
factor 2 near the peak of the cross section $d\sigma/d\eta$. Note
that our results coincide with the results of the Ref.\cite{21} in
case of the JB parameterization.

The $p_T$ spectrum of $D^\star$ meson in photoproduction at
$|\eta| < 1.5$ and $130 < W < 280$ GeV are shown in Fig.4. The
theoretical curve in the $k_T$-factorization approach obtained
using the JB parameterization describes data well at the all
$p_T$. The curves obtained using KMR or JS parameterizations
coincide with the data only at the $p_T\le 8$ GeV. The typical
value of the $K$-factor in case of LO parton model  calculation is
equal 2 at the large $p_T$.

Our results show that the introducing of a gluon transverse
momentum $k_T$ in the framework of the $k_T$-factorization
approach increase the $D^\star$ meson photoproduction cross
section at the large $p_T$ in case of the JS parameterization, as
it is predicted for the $J/\Psi$ photoproduction \cite{22,23} (see
next part of the paper) and at the all relevant values of the
$p_T$ in case of KMR or JB parameterizations.

The dependence of the $D^\star$ meson production cross section on
a total photon-proton center-of-mass energy W is shown in Fig.5.
Nowadays the experimental data for $d\sigma /dW$ are absent. We
see that the difference between the results obtained with the
various parameterizations of a unintegrated gluon distribution
function is about 50\%. As well in the case of the $\eta$-spectra.
At the all $p_{Tmin}$ the cross section calculated in the parton
model is smaller than predictions obtained in the
$k_T$-factorization approach.

The main uncertainties of our calculation come from the choice of
a $c$-quark mass in the partonic matrix elements (10),(11) and
(13), and from the choice of a parameter $\epsilon$ in the
Peterson fragmentation function $D_{c \to
D^\star}(z,\mu^2)$.

We see that agreement between the results obtained in the
$k_T$-factorization approach and the data on the $D^*$ meson
photoproduction is better at the large $p_T$. This fact shows that
in the $k_T$-factorization approach the next to leading order
corrections are taken into account effectively.

\section{$J/\psi$ photoproduction in LO QCD}

It is well known that in the processes of $J/\psi$ meson
photoproduction on protons at high energies the photon-gluon
fusion partonic subprocess dominates \cite{24}. In the framework
of the general factorization approach of QCD the $J/\psi$
photoproduction cross section depends on the gluon distribution
function in a proton, the hard amplitude of $c\bar c$-pair
production as well as the mechanism of a creation colorless final
state with quantum numbers of the $J/\psi$ meson. In such a way,
we suppose that the soft interactions in the initial state are
described by introducing a gluon distribution function, the hard
partonic amplitude is calculated using perturbative theory of QCD
at order in $\alpha_s(\mu^2)$, where $\mu\sim m_c$, and the soft
process of the $c\bar c$-pair transition into the $J/\psi$ meson
is described in nonrelativistic approximation using series in the
small parameters $\alpha_s$ and $v$ (relative velocity of the
quarks in the $J/\psi$ meson). As is said in nonrelativistic QCD
(NRQCD) \cite{25}, there are color singlet mechanism, in which the
$c\bar c$-pair is hardly produced in the color singlet state, and
color octet mechanism, in which the $c\bar c$-pair is produced in
the color octet state and at a long distance it transforms into a
final color singlet state in the soft process. However, as it was
shown in papers \cite{23,26}, the data from the DESY ep-collider
\cite{27} in the wide region of $p_T$ and $z$ may be described
well in the framework of the color singlet model and the color
octet contribution is not needed. Based on the above mentioned
result we will take into account in our analysis only the color
singlet model contribution in the $J/\psi$ meson photoproduction
\cite{24}. We consider here the role of a proton gluon
distribution function in the $J/\psi$ photoproduction in the
framework of the conventional parton model as well as in the
framework of the $k_T$-factorization approach \cite{3,4,5}.

There are six Feynman diagrams (Fig.6) which describe the partonic
process $\gamma g \to J/\psi g$ at the leading order in $\alpha_s$
and $\alpha$. In the framework of the color singlet model and
nonrelativistic approximation the production of the $J/\psi$ meson
is considered as the production of a quark-antiquark system in the
color singlet state with orbital momentum $L=0$ and spin momentum
$S=1$. The binding energy and relative momentum of quarks in the
$J/\psi$ are neglected. In such a way $M=2 m_c$ and $p_c=p_{\bar
c}=\displaystyle{\frac{p}{2}}$, where $p$ is the 4-momentum of the
$J/\psi$,  $p_c$ and $p_{\bar c}$ are 4-momenta of quark and
antiquark. Taking into account the formalism of the projection
operator \cite{28} the amplitude of the process $\gamma g\to
J/\psi g$ may be obtained from the amplitude of the process
$\gamma g\to \bar c c g$ after replacement:
\begin{equation}
V^i(p_{\bar c})\bar U^j(p_c) \rightarrow
\frac{\Psi(0)}{2\sqrt{M}}\hat \varepsilon (p)(\hat
p+M)\frac{\delta^{ij}}{\sqrt{3}}\mbox{,}
\end{equation}
where $\hat \varepsilon (p)=\varepsilon_{\mu}(p)\gamma^{\mu}$,
$\varepsilon_{\mu}(p)$ is a 4-vector of the $J/\psi$ polarization,
$\displaystyle{\frac{\delta^{ij}}{\sqrt{3}}}$ is the color factor,
$\Psi(0)$ is the nonrelativistic meson wave function at the
origin. The matrix elements of the process $\gamma g^\star\to
J/\psi g$ may be presented as follows:
\begin{equation}
M_i=KC^{ab}\varepsilon_{\alpha}(q_1)\varepsilon^a_{\mu}(q)
\varepsilon^b_{\beta}(q_2)\varepsilon_{\nu}(p)
M_i^{\alpha\beta\mu\nu}\mbox{,}
\end{equation}
\begin{equation}
M_1^{\alpha\beta\mu\nu}= \mbox{Tr}\left[\gamma^{\nu}(\hat
p+M)\gamma^{\alpha}\frac{\hat p_c-\hat
q_1+m_c}{(p_c-q_1)^2-m_c^2}\gamma^{\mu}\frac{-\hat p_{\bar c}-\hat
q_2+m_c}{(p_{\bar c}+q_2)^2-m_c^2}\gamma^{\beta}\right]\mbox{,}
\end{equation}
\begin{equation}
M_2^{\alpha\beta\mu\nu}= \mbox{Tr}\left[\gamma^{\nu}(\hat
p+M)\gamma^{\beta}\frac{\hat p_c+\hat
q_2+m_c}{(p_c+q_2)^2-m_c^2}\gamma^{\alpha}\frac{\hat k -\hat
p_{\bar c}+m_c}{(q-p_{\bar c})^2-m_c^2}\gamma^{\mu}\right]\mbox{,}
\end{equation}
\begin{equation}
M_3^{\alpha\beta\mu\nu}= \mbox{Tr}\left[\gamma^{\nu}(\hat
p+M)\gamma^{\alpha}\frac{\hat p_c-\hat
q_1+m_c}{(p_c-q_1)^2-m_c^2}\gamma^{\beta}\frac{\hat k-\hat p_{\bar
c}+m_c}{(q-p_{\bar c})^2-m_c^2}\gamma^{\mu}\right]\mbox{,}
\end{equation}
\begin{equation}
M_4^{\alpha\beta\mu\nu}= \mbox{Tr}\left[\gamma^{\nu}(\hat
p+M)\gamma^{\mu}\frac{\hat p_c-\hat
k+m_c}{(p_c-q)^2-m_c^2}\gamma^{\alpha}\frac{-\hat p_{\bar c}-\hat
q_2+m_c}{(q_2+p_{\bar c})^2-m_c^2}\gamma^{\beta}\right]\mbox{,}
\end{equation}
\begin{equation}
M_5^{\alpha\beta\mu\nu}= \mbox{Tr}\left[\gamma^{\nu}(\hat
p+M)\gamma^{\beta}\frac{\hat p_c+\hat
q_2+m_c}{(p_c+q_2)^2-m_c^2}\gamma^{\mu}\frac{\hat q_1 -\hat
p_{\bar c}+m_c}{(q_1-p_{\bar
c})^2-m_c^2}\gamma^{\alpha}\right]\mbox{,}
\end{equation}
\begin{equation}
M_6^{\alpha\beta\mu\nu}= \mbox{Tr}\left[\gamma^{\nu}(\hat
p+M)\gamma^{\mu}\frac{\hat p_c-\hat
k+m_c}{(p_c-q)^2-m_c^2}\gamma^{\beta}\frac{\hat q_1-\hat p_{\bar
c}+m_c}{(q_1-p_{\bar c})^2-m_c^2}\gamma^{\alpha}\right]\mbox{,}
\end{equation}
where  $q_1$ is the 4-momentum of the photon,  $q$ is the
4-momentum of the initial gluon, $q_2$ is the  4-momentum of the
final gluon,
$$
K=e_ceg_s^2\frac{\Psi(0)}{2\sqrt{M}}\mbox{,}\qquad
C^{ab}=\frac{1}{\sqrt{3}}\mbox{Tr}[T^aT^b]\mbox{,}\quad
e_c=\frac{2}{3}\mbox{,}\quad e=\sqrt{4\pi\alpha}\mbox{,}\quad
g_s=\sqrt{4\pi\alpha_s}\mbox{.}
$$
The summation on the photon, the $J/\psi$ meson and final gluon
polarizations is carried out by covariant formulae:
\begin{eqnarray}
&&\sum_{spin}\varepsilon_{\alpha}(q_1)\varepsilon_{\beta}(q_1)
=-g_{\alpha\beta},\\
&&\sum_{spin}\varepsilon_{\alpha}(q_2)\varepsilon_{\beta}(q_2)
=-g_{\alpha\beta},\\
&&\sum_{spin}\varepsilon_{\mu}(p)\varepsilon_{\nu}(p)
=-g_{\mu\nu}+\frac{p_{\mu}p_{\nu}}{M^2}.
\end{eqnarray}
In case of the initial BFKL gluon we use the prescription (5).
%\begin{equation}
%\sum_{spin}\varepsilon_{\mu}(q)\varepsilon_{\nu}(q)=
%\frac{q_{T\mu}q_{T\nu}}{{\bf k}^2_{T}}.
%\end{equation}
For studing  $J/\psi$ polarized photoproduction we introduce the
4-vector of the longitudinal polarization as follows:
\begin{equation}
\varepsilon_L^{\mu}(p)=\frac{p^{\mu}}{M}-\frac{Mp_N^{\mu}}{(pp_N)}.
\end{equation}
In the high energy limit of $s=2(q_1p_N)>>M^2$ the polarization
4-vector satisfies usual conditions
$(\varepsilon_L\varepsilon_L)=-1$, $(\varepsilon_L p)=0$.

Traditionally for a description of charmonium photoproduction
processes the invariant variable $z=(pp_N)/(q_1p_N)$ is used. In
the rest frame of the proton one has $z=E_{\psi}/E_{\gamma}$. In
the $k_T$-factorization approach the differential on $p_{T}$ and
$z$ cross section of the $J/\psi$ photoproduction may be written
as follows:
\begin{equation}
\frac{d\sigma(\gamma p\to J/\Psi
X)}{dp^2_{T}dz}=\frac{1}{z(1-z)}\int_0^{2\pi}
\frac{d\varphi}{2\pi}\int_0^{\mu^2} d{\bf k}^2_{T}\Phi (x,{\bf
k}^2_{T},{\mu}^2) \frac{\overline{|M|^2}}{16\pi(xs)^2}\mbox{.}
\end{equation}

The analytical calculation of the $\overline{|M|^2}$ is performed
with help of REDUCE package and results are saved in the FORTRAN
codes as a function of $\hat s=(q_1+q)^2$, $\hat t=(p-q_1)^2$,
$\hat u=(p-q)^2$, ${\bf p}_T^2$, ${\bf k}^2_{T}$ and  $\cos
(\varphi)$. We directly have tested that
\begin{equation}
\lim_{{{\bf k}^2_{T}\to 0}}\int_0^{2\pi}\frac{d\varphi}{2\pi}
\overline{|M|^2}= \overline{|M_{PM}|^2}\mbox{,}
\end{equation}
where ${\bf p}_T^2= \displaystyle{\frac{\hat t \hat u}{\hat s}}$
in the $\overline{|M|^2}$ and $\overline{|M_{PM}|^2}$ is the
square of the amplitude in the conventional parton model
\cite{24}. In the limit of ${\bf k}^2_{T}=0$ from formula (27) it
is easy to find the differential cross section in the parton
model, too:
\begin{equation}
\frac{d\sigma^{PM}(\gamma p\to J/\Psi X)}{dp^2_{T}dz}=
\frac{\overline{|M_{PM}|^2}xG(x,\mu^2)}{16\pi(xs)^2z(1-z)}\mbox{.}
\end{equation}
However, making calculations in the parton model we use formula
(27), where integration over ${\bf k}^2_{T}$ and $\varphi$ is
performed numerically, instead of (29). This method fixes the
common normalization factor for both approaches and gives a direct
opportunity to study effects connected with virtuality of the
initial BFKL gluon in the partonic amplitude.

\section{$J/\psi$ photoproduction at HERA}

After we fixed the selection of the gluon distribution functions
$G(x,\mu^2)$ or $\Phi (x,{\bf k}^2_{T},{\mu}^2)$ there are two
parameters only, which values determine the common normalization
factor of the cross section under consideration: $\Psi(0)$ and
$m_c$. The value of the $J/\psi$ meson wave function at the origin
may be calculated in a potential model or obtained from
experimental well known decay width $\Gamma(J/\psi\to
\mu^+\mu^-)$. In our calculation we used the following choice
$|\Psi(0)|^2=0.064$ GeV$^3$ which corresponds to NRQCD coefficient
$\langle O^{J/\psi},{\bf 1}^3S_1\rangle =1.12$ GeV$^3$ as the same
as in Ref. \cite{26}. Note, that this value is a little smaller
(30 \%) than the value which was used in our paper \cite{23}.
Concerning a charmed quark mass, the situation is not clear up to
the end. From one hand, in the nonrelativistic approximation one
has $m_c=\displaystyle{\frac{M}{2}}$, but there are many examples
of taking smaller value of a $c$-quark mass in the amplitude of a
hard process, for example, $m_c=1.4$ GeV. Taking into
consideration above mentioned we perform calculations at $m_c=1.5$
GeV. The cinematic region under consideration is determined by the
following conditions: $Q^2<1$ GeV$^2$, $60<W<240$ GeV, $0.3<z<0.9$
and $p_T>1$ GeV, which correspond to the H1 Collaboration data
\cite{29}. We assume that the contribution of the color octet
mechanism is large at the $z>0.9$ only. In the region of the small
values of the  $z<0.2$ the contribution of the resolved photon
processes \cite{30} as well as the charm excitation processes
\cite{31} may be large, too. All of these contributions are not in
our consideration.

Figs. 7--10 show our results which were obtained as in the
conventional parton model as well as in the $k_T$-factorization
approach with the different parameterizations of the unintegrated
gluon distribution function. The dependence of the results on
selection of a hard scale parameter $\mu$ is much less than the
dependence on selection of a $c$-quark mass and selection of a
parameterization. We put $\mu^2=M^2+ \bf p_T^2$ in a gluon
distribution function and in a running constant $\alpha_s(\mu^2)$.

 The count of a transverse momentum of the BFKL gluons in the
$k_T$-factorization approach results in a flattening of the
$p_T$-spectrum of the $J/\psi$ as contrasted by predictions of the
parton model. For the first time this effect was indicated in the
Ref. \cite{22}, and later in the Ref. \cite{23}. Fig. 7 shows the
result of our calculation for the $p_T$ spectrum of the $J/\psi$
mesons. Using the $k_T$-factorization approach we have obtained
the harder $p_T$-spectrum of the  $J/\psi$ than has been predicted
in the LO parton model. It is visible that at large values of
$p_T$ only the $k_T$-factorization approach gives correct
description of the data \cite{29}. However, it is impossible to
consider this visible effect as a direct indication on nontrivial
developments of the small-x physics. In the article \cite{26} was
shown that the calculation in the NLO approximation using
colliniar parton model gives a harder $p_T$ spectrum of the
$J/\psi$ meson, too, which will agree with the data at the large
$p_T$.

In the $k_T$-factorization approach JB parameterization \cite{8}
gives $p_T$-spectrum, which is very close to experimental data.
From the another hand in the case of JS parameterization \cite{9}
the additional  $K$-factor approximately equal 2 is needed.

The $z$ spectra are shown in Fig.8 at the various choice of the
$p_T$ cut: $p_T > 2$, $4$ and $6$ GeV, correspondingly. The
relation between the theoretical predictions and experimental data
is the same as in Fig.7. The $k_T$ factorization approach give
more correct description of the data especially at large value of
$z$ where the curve obtained in the collinear parton model tends
to zero.

Fig.9 shows the dependence of the total $J/\psi$ photoproduction
cross section on $W$ at $0.3<z<0.8$ and $p_T>1$ GeV. The shape of
this dependence agrees well with the result obtained using JS
parameterization \cite{9} or KMR \cite{10} parameterization.
However, the predicted absolute value of the cross section
$\sigma_{\gamma p}$ is smaller by factor 2 than obtained data
\cite{29}. The results of calculation using JB \cite{8} or GRV
\cite{7} parameterizations are larger and coincide with the data
\cite{29} better.

As it was mentioned above, the main difference between the
$k_T$-factorization approach and the conventional parton model is
nontrivial polarization of the BFKL gluon. It is obvious, that
such a spin condition of the initial gluon should result in
observed spin effects during the birth of the polarized $J/\psi$
meson. We have performed calculations for the spin parameter
$\alpha$ as a function $z$ or $p_T$ in the conventional parton
model and in the $k_T$-factorization approach:
\begin{equation}
\alpha (z)={\frac {\displaystyle\frac{d\sigma_{tot}}{dz}
-3\frac{d\sigma_L}{dz}}
{\displaystyle\frac{d\sigma_{tot}}{dz}+\frac{d\sigma_L}{dz}}},\qquad
\alpha (p_T)={\frac { \displaystyle\frac{d\sigma_{tot}}{dp_{T}}
-3\frac{d\sigma_L}{dp_T}}
{\displaystyle\frac{d\sigma_{tot}}{dp_T}+\frac{d\sigma_L}{dp_T}}}
\end{equation}
Here  $\sigma_{tot}=\sigma_L+\sigma_T$ is the total $J/\psi$
production cross section, $\sigma_L$ is the production cross
section for the longitudinal polarized $J/\psi$ mesons, $\sigma_T$
is the production cross section for the transverse polarized
$J/\psi$ mesons. The parameter $\alpha$ controls the angle
distribution for leptons in the decay $J/\psi \to l^+l^-$ in the
$J/\psi$ meson rest frame:
\begin{equation}
\frac{d\Gamma}{d \cos(\theta)}\sim 1+\alpha\cos^2(\theta).
\end{equation}

The theoretical results for the parameter $\alpha(z)$ are very
close to each other irrespective of the choice of an approach or a
gluon parameterization \cite{23}.

For the parameter $\alpha (p_T)$ we have found strongly opposite
predictions in the parton model and in the $k_T$-factorization
approach, as it is visible in Fig. 10. The parton model predicts
that  $J/\psi$ mesons should have transverse polarizations at the
large $p_T$ ($\alpha(p_T)=0.6$ at the $p_T=6$ GeV), but
$k_T$-factorization approach predicts that $J/\psi$ mesons should
be longitudinally polarized ( $\alpha(p_T)=-0.4$ at the $p_T=6$
GeV). The experimental points lie in the range $0 < p_T < 5$ GeV
and they have the large errors. However, it is visible that
$\alpha(p_T)$ decrease as $p_T$ changes from 1 to 5 GeV. This fact
coincide with theoretical prediction obtained in the
$k_T$-factorization approach. Nowadays, a result of the NLO parton
model calculation in the case of the polarized $J/\psi$ meson
photoproduction is unknown. It should be an interesting subject of
future investigations. If the count of the NLO corrections will
not change predictions of the LO parton model for $\alpha (p_T)$,
the experimental measurement of this spin effect will be a direct
signal about BFKL gluon dynamics.

Nowadays, the experimental data on $J/\psi$ polarization in
photoproduction at large $p_T$ are absent. However there are
similar data from CDF Collaboration \cite{32}, where $J/\psi$ and
$\psi'$ $p_T$-spectra and polarizations have been measured.
Opposite the case of $J/\psi$ photoproduction, the hadroproduction
data needs to take into account the large color-octet contribution
in order to explain $J/\psi$ and $\psi'$ production at Tevatron in
the conventional collinear parton model. The relative weight of
color-octet contribution may be smaller if we use
$k_T$-factorization approach, as was shown recently in
\cite{33,34,35,36}. The predicted using collinear parton model
transverse polarization of $J/\psi$ at large $p_T$ is not
supported by the CDF data, which can be roughly explained by the
$k_T$-factorization approach \cite{34}. In conclusion, the number
of theoretical uncertainties in the case of $J/\psi$ meson
hadroproduction is much more than in the case of photoproduction
and they need more complicated investigation, which is why the
future experimental analysis of $J/\psi$ photoproduction at THERA
will be clean check of the collinear parton model and the
$k_T$-factorization approach.

\section{Acknowledgements}
The authors would like to thank M.~Ivanov and S.~Nedelko for kind
hospitality during International School "Heavy quark physics -
2002" in Dubna, S.~Baranov, A.~Lipatov and O.~Teryaev for
discussions on the $k_T$-factorization approach of QCD and H.~Jung
for the valuable information on unintegrated gluon distribution
functions. This work has been supported in part by the Russian
Foundation for Basic Research under Grant 02-02-16253.

%\begin{thebibliography}{99}

\begin{figure}[ht]
\begin{center}
\includegraphics[width=1.0\textwidth, clip=]{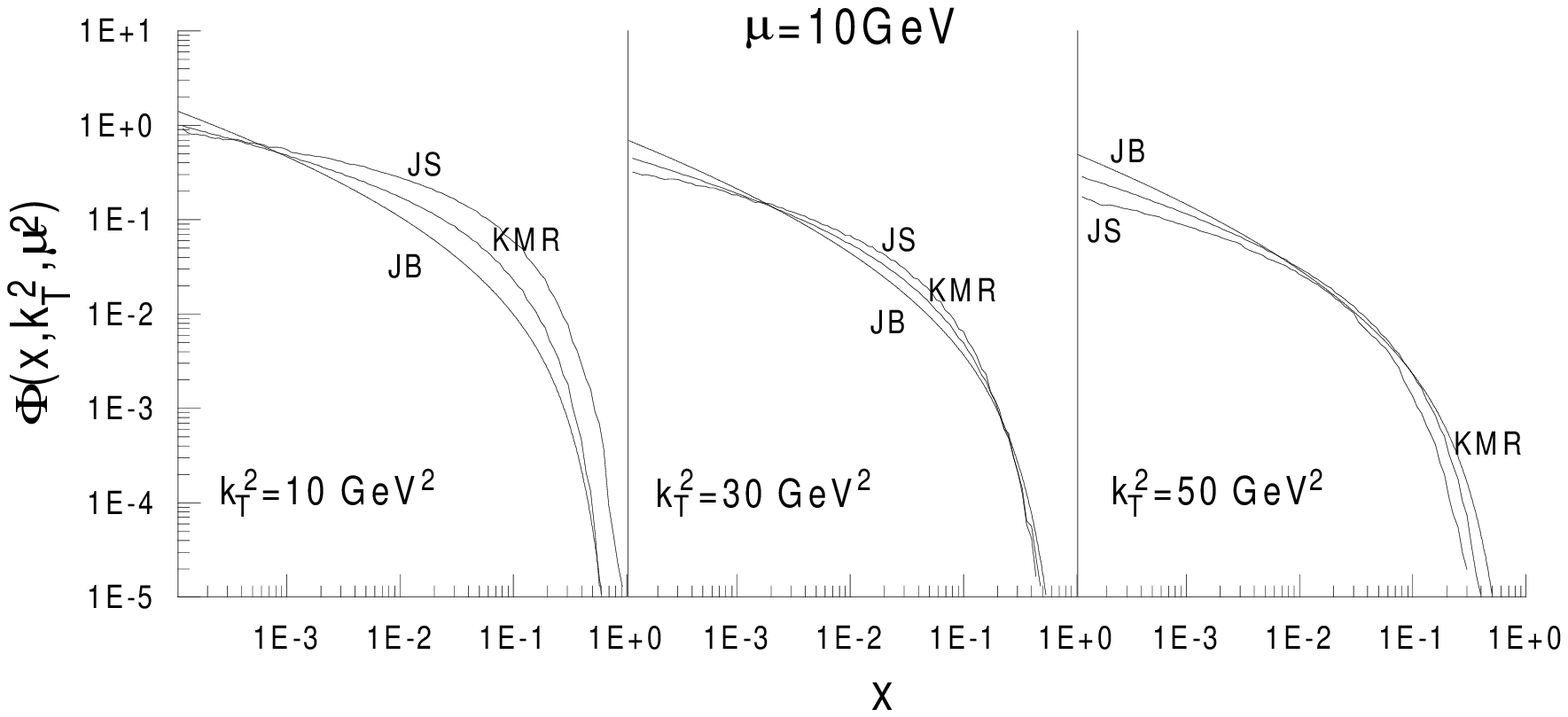}
\end{center}
\caption[]{The unintegrated gluon distribution function
$\Phi(x,\vec k_T^2,\mu^2)$ versus $x$ at the fixed values of $\mu$
and $k_T^2$.} \label{eps1}
\end{figure}
\begin{figure}[ht]
\begin{center}
\includegraphics[width=1.0\textwidth, clip=]{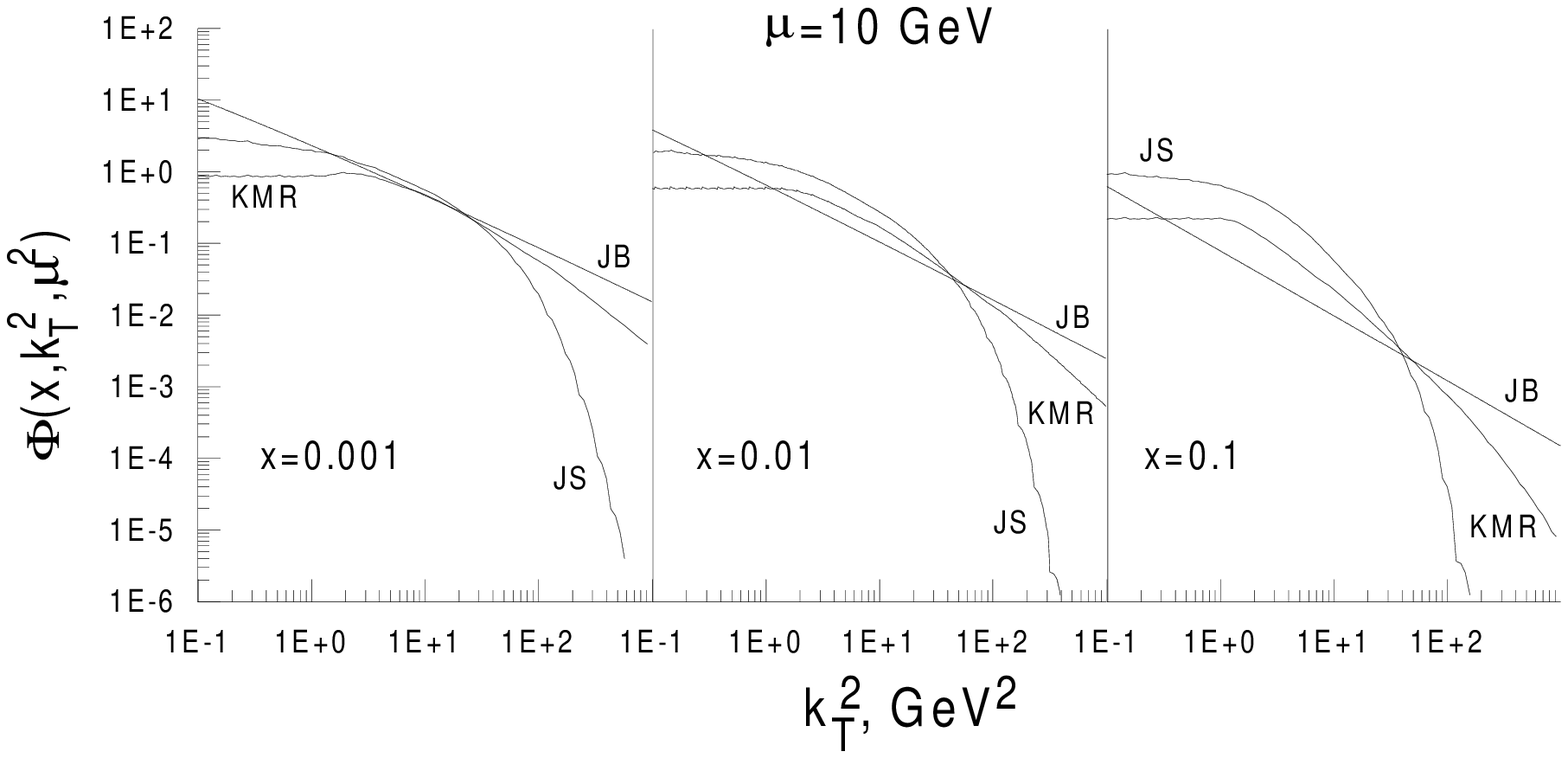}
\end{center}
\caption[]{The unintegrated gluon distribution function
$\Phi(x,\vec k_T^2,\mu^2)$ versus $k_T^2$ at the fixed values of
$\mu$ and $x$.} \label{eps2}
\end{figure}
\begin{figure}[ht]
\begin{center}
\includegraphics[width=.8\textwidth, clip=]{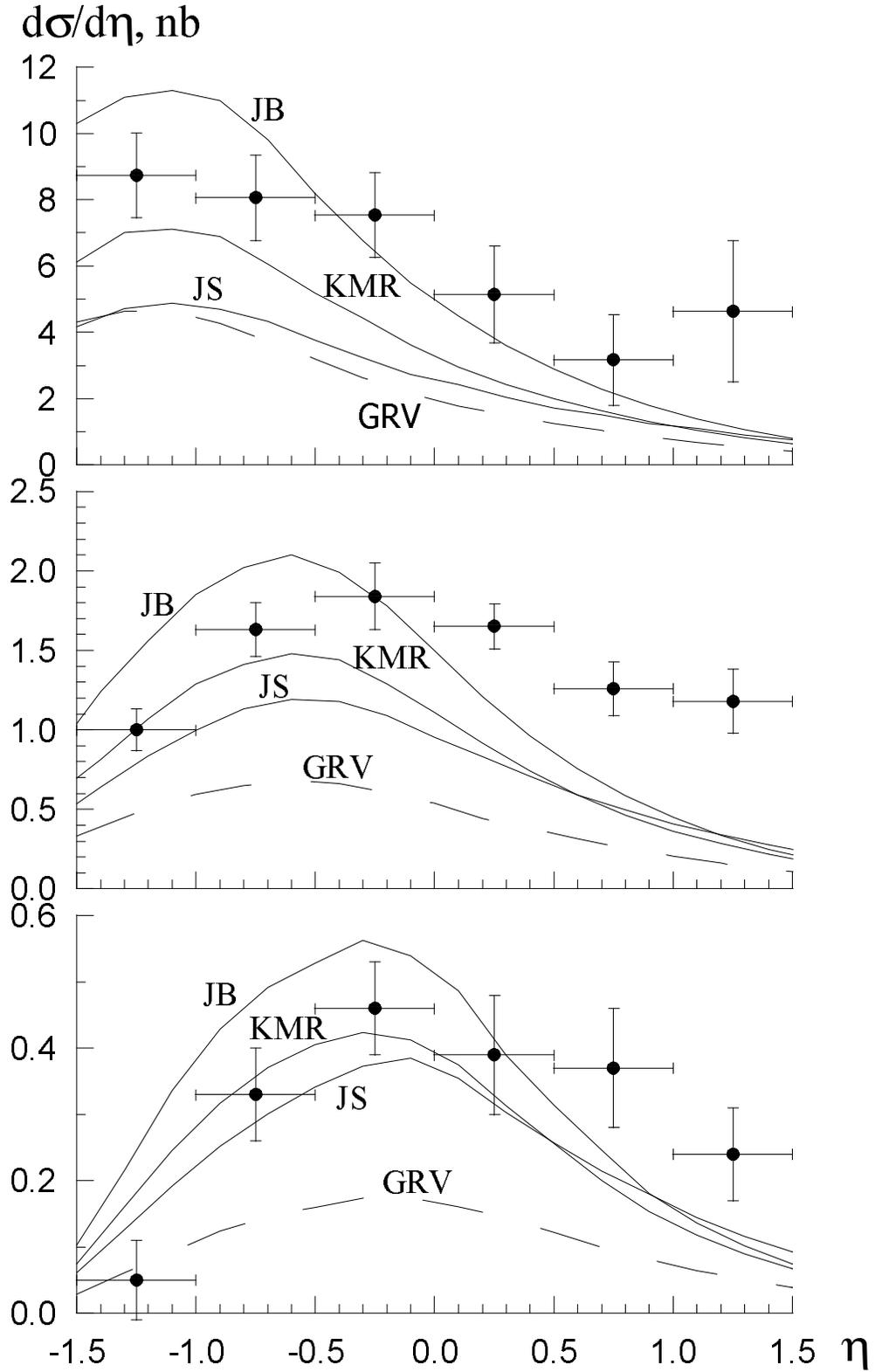}
\end{center}
\caption[]{The $\eta$ spectra of the $D^\star$ meson at the
various cut on a transverse momentum ($p_T>2, 4, 6$ GeV,
correspondingly from up to down) and $130 < W < 280$ GeV.}
\label{eps3}
\end{figure}
\begin{figure}[ht]
\begin{center}
\includegraphics[width=0.8\textwidth, clip=]{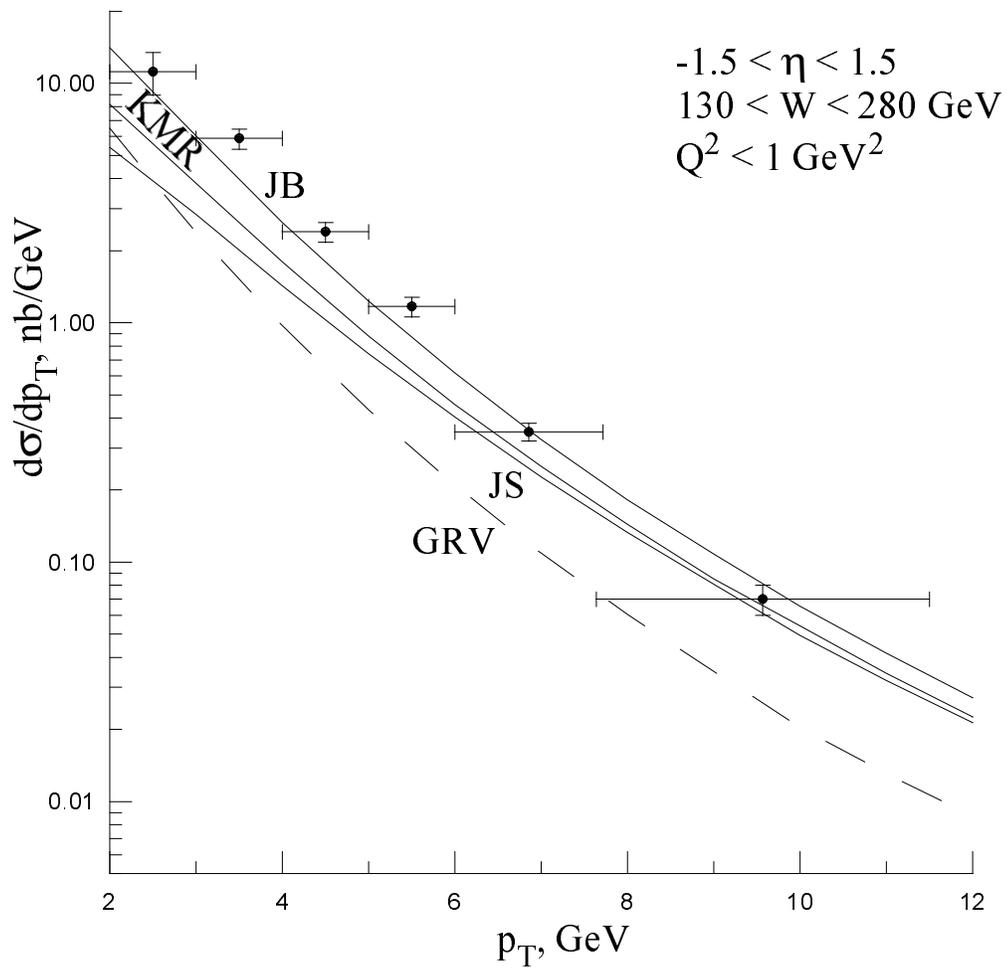}
\end{center}
\caption[]{The $p_T$-spectrum of $D^\star$ meson.} \label{eps4}
\end{figure}

\begin{figure}[ht]
\begin{center}
\includegraphics[width=.8\textwidth, clip=]{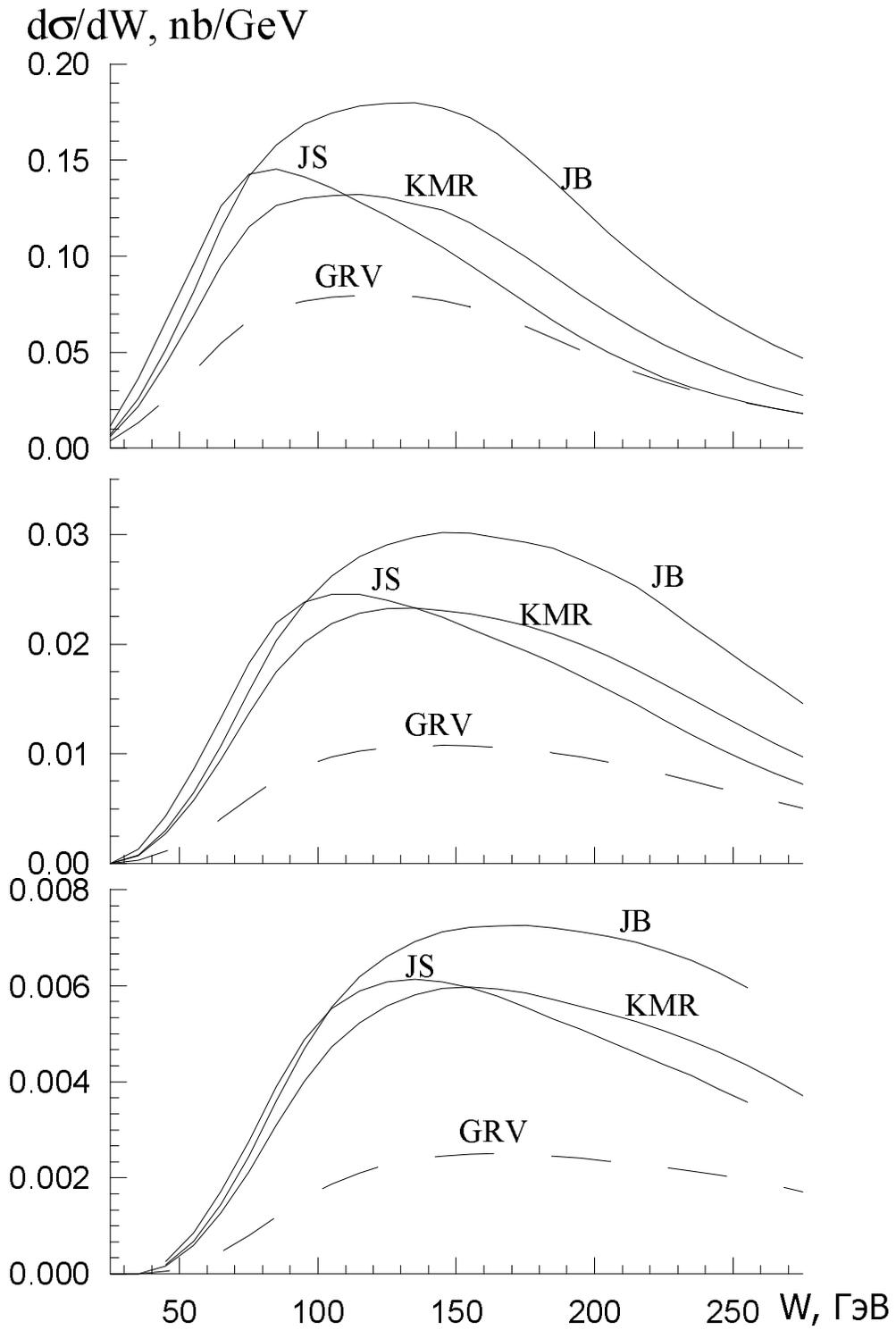}
\end{center}
\caption[]{The theoretical predictions for the $W$-spectra at the
various cut on the $D^\star$ meson transverse momentum ($p_T>2, 4,
6$ GeV, correspondingly from up to down) and $|\eta|<1.5$. }
\label{eps5}
\end{figure}
\begin{figure}[ht]
\begin{center}
\includegraphics[width=.6\textwidth, clip=]{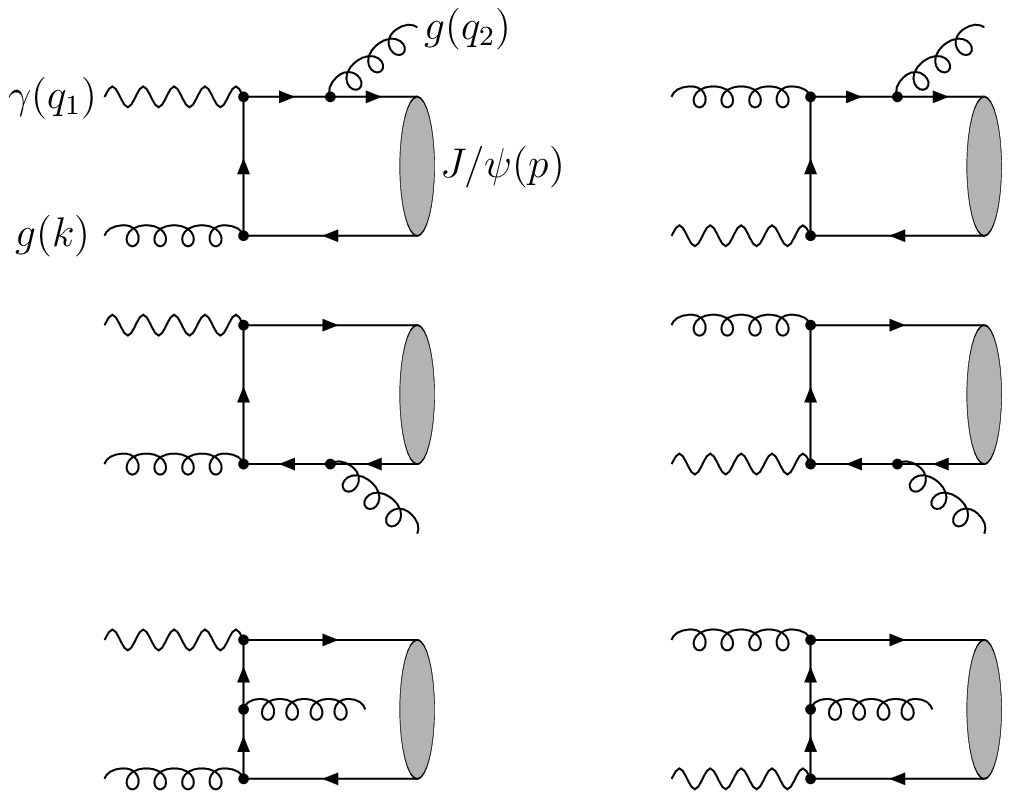}
\end{center}
\caption[]{Diagrams used for description partonic process
$\gamma+g\to J/\psi+g$. } \label{eps6}
\end{figure}

\begin{figure}[ht]
\begin{center}
\includegraphics[width=.8\textwidth, clip=]{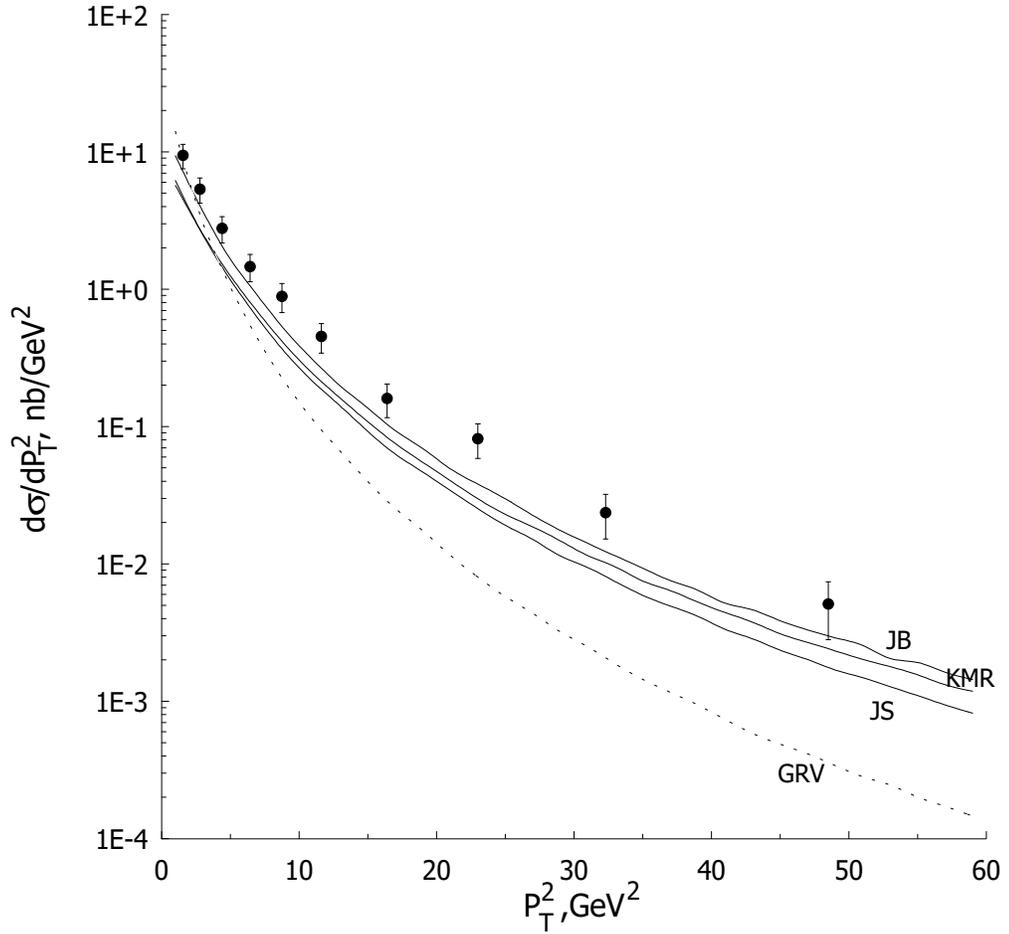}
\end{center}
\caption[]{The  $J/\psi$ spectrum on $p^2_{T}$ at the $60<W<240$
GeV and $0.3<z<0.9$. } \label{eps7}
\end{figure}
\begin{figure}[ht]
\begin{center}
\includegraphics[width=.8\textwidth, clip=]{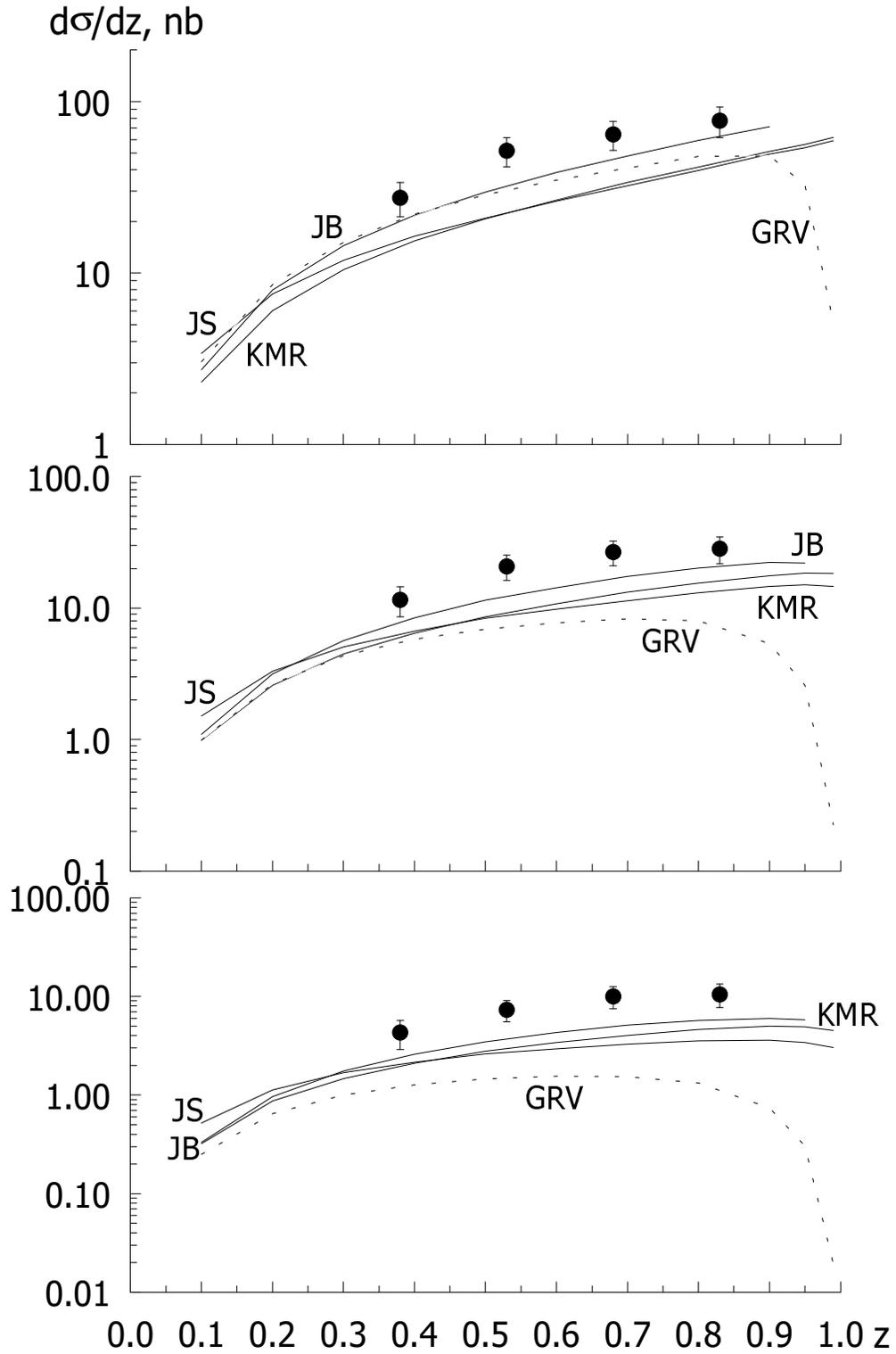}
\end{center}
\caption[]{The
 $J/\psi$ spectrum on $z$ at the $60<W<240$ GeV (
$p_{T}>1,2,3$ GeV, correspondingly from up to down).  }
\label{eps8}
\end{figure}

%\vspace{50mm}

\begin{figure}[ht]
\begin{center}
\includegraphics[width=.6\textwidth, clip=]{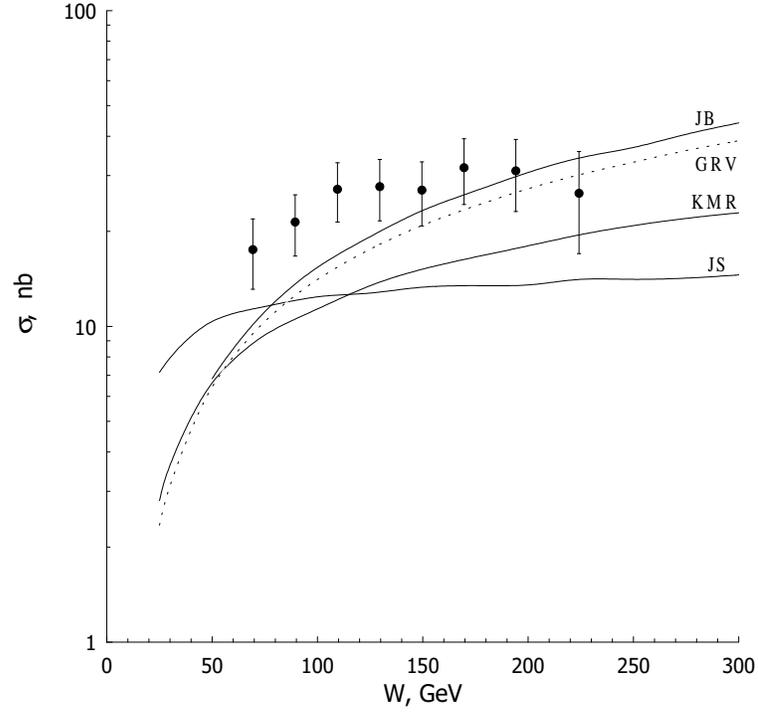}
\end{center}
\caption[]{The total $J/\psi$ photoproduction cross section versus
$W$ at the $0.3<z<0.8$ and $p_T>1$ GeV.} \label{eps9}
\end{figure}

\begin{figure}[ht]
\begin{center}
\includegraphics[width=.6\textwidth, clip=]{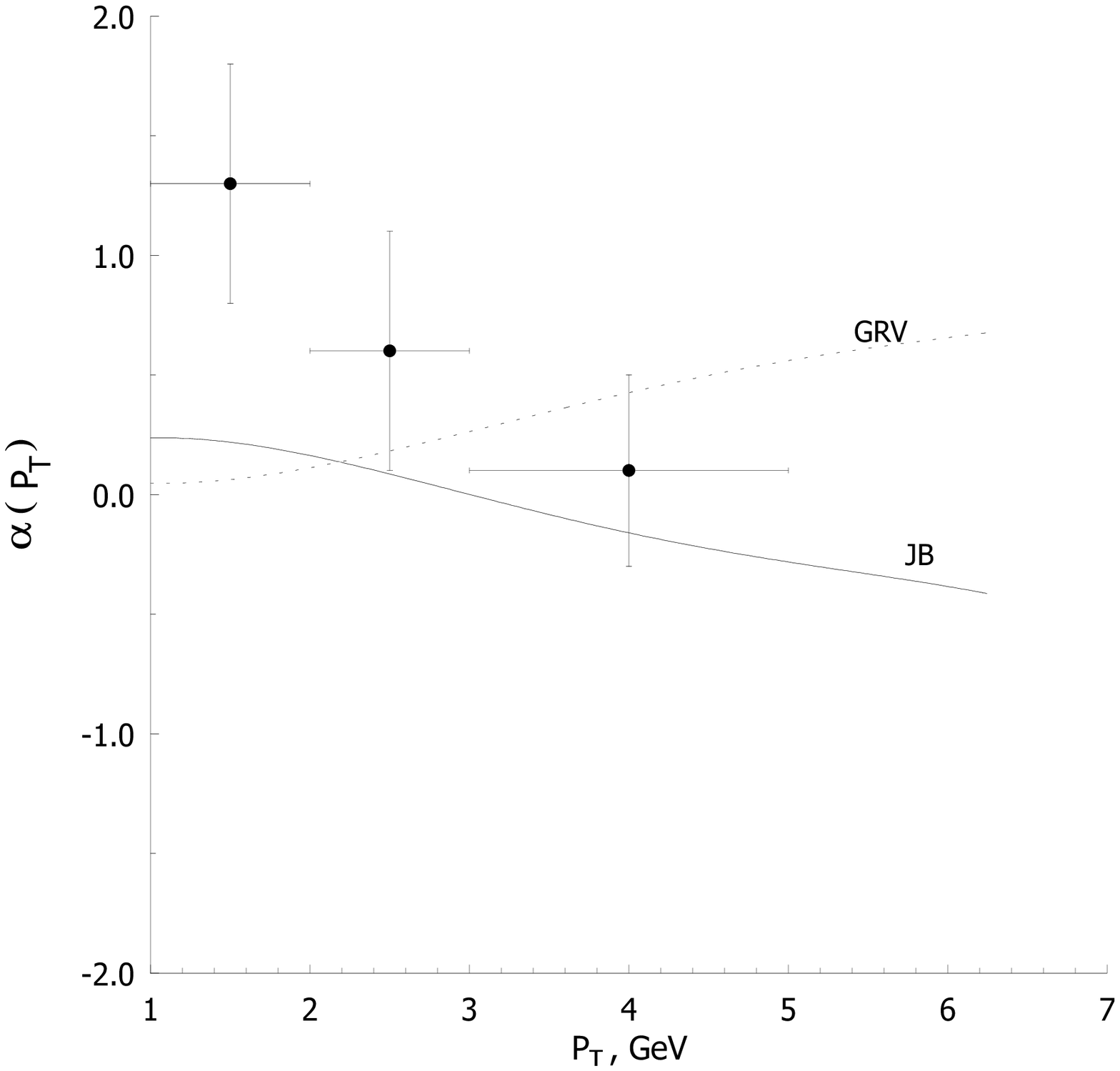}
\end{center}
\caption[]{Parameter $\alpha$ as a function of $p_T$ at the $0.3 <
z < 0.9$, $60 < W < 240$ GeV.} \label{eps10}
\end{figure}

\begin{thebibliography}{8.}
\addcontentsline{toc}{section}{References}
\bibitem{1}G.~Sterman et al., Rev. Mod. Phys.  {59} (1995) 158.
\bibitem{2}V.N.~Gribov and L.N.~Lipatov, Sov. J. Nucl. Phys.  {15} (1972) 438
; \newline Yu.A.~Dokshitser, Sov. Phys. JETP.  {46} (1977) 641
;\newline G.~Altarelli and G.~Parisi, Nucl. Phys.  {B126} (1977)
298.
\bibitem{3}L.V.~Gribov, E.M.~Levin, M.G. Ryskin, Phys. Rep.
 {100} (1983) 1.
\bibitem{4}J.C.~Collins and R.K.~Ellis, Nucl. Phys.  {360} (1991)
3.
\bibitem{5}S.~Catani,M.~Ciafoloni and F.~Hautmann, Nucl. Phys.  {B366} (1991)
135.
\bibitem{6}E.~Kuraev, L.~Lipatov, V.~Fadin, Sov. Phys. JETP  {44} (1976) 443
;\newline Y.~Balitskii and L.~Lipatov, Sov. J. Nucl. Phys.
 {28} (1978) 822.
\bibitem{7}M.~Gluck, E.~Reya and A.~Vogt, Z. Phys.  {C67} (1995)
433.
\bibitem{8}J.~Bluemlein, DESY 95-121.
\bibitem{9}H.~Jung, G.~Salam, Eur. Phys. J.  {C19} (2001) 351.
\bibitem{10}M.A.~Kimber, A.D.~Martin and M.G.~Ryskin, Phys. Rev.  {D63} (2001)
114027.
\bibitem{11}J.~Breitweg at al.[ZEUS Coll.], Eur. Phys. J.  {C6} 67 (1999)
;\newline J.~Breitweg at al.[ZEUS Coll.], Phys. Lett.  {B481}
(2000) 213.
\bibitem{12}C.~Adloff et al.[H1 Coll.], Nucl. Phys.  {B545} (1999)
21.
\bibitem{13}C.~Peterson et al.,Phys. Rev.  {D27} (1983) 105.
\bibitem{14}A.V.~Berezhnoy, V.V.~Kiselev and A.K.~Likhoded, hep-ph/9901333,
; \newline A.V.~Berezhnoy, V.V.~Kiselev and A.K.~Likhoded,
hep-ph/9905555.
\bibitem{15}R.~Akers et al.[OPAL Coll.], Z. Phys.  {C67} (1995)
27.
\bibitem{16}P.~Nason and C.~Oleari, Phys. Lett.  {B447} (1999)
327.
\bibitem{17}B.~Kniehe et al., Z. Phys.  {C76}, (1997) 689;\newline
J.~Binnewics et al., Phys. Rev.  {D58} (1998) 014014.
\bibitem{18}V.A.~Saleev and N.P.~Zotov, Mod. Phys. Lett.  {A11} (1996)
25.
\bibitem{19}V.M.~Budnev et al., Phys. Rep.  {15} (1974) 181.
\bibitem{20}S.~Frixione et al., Phys. Lett.  {B319} (1993) 339.
\bibitem{21}S.~Baranov, N.~Zotov, Phys. Lett.  {B458} (1999) 389.
\bibitem{22}V.A.~Saleev, N.P.~Zotov, Mod. Phys. Lett.  {A9} (1994) 151,
1517.
\bibitem{23}V.A.~Saleev, Phys. Rev.  {D65} (2002) 054041.
%%%%%%%%%%%%%%%%%%%%%%%%%%
\bibitem{24}E.L.~Berger and D.~Jones, Phys. Rev.  {D23} (1981)
1521 ; \newline R.~Baier and R.~Ruckl, Phys. Lett.  {B102} (1981)
364 ; \newline S.S.~Gershtein, A.K.~Likhoded, S.R.~Slabospiskii,
Sov. J. Nucl. Phys.  {34}, (1981) 128.
\bibitem{25}G.T.~Bodwin, E.~Braaten, G.P.~Lepage, Phys. Rev
 {D51} (1995) 1125.
\bibitem{26}M.~Kramer, Nucl. Phys.  {B459} (1996) 3.
\bibitem{27}Aid et al.[H1 Coll.], Nucl. Phys.  {B472} (1996) 32
; \newline J.~Breitweg et al.[ZEUS Coll.], Z. Phys.  {C76} (1997)
599.
\bibitem{28}B.~Guberina et al., Nucl. Phys.  {B174} (1980) 317.
\bibitem{29}C.~Adloff et al.[H1 Coll.], DESY 02-059 (2002).
\bibitem{30}H.~Jung, G.A.~Schuler and J.~Terron, DESY 92-028
(1992).
\bibitem{31}V.A.~Saleev, Mod. Phys. Lett.  {A9} (1994) 1083.
\bibitem{32}T.~Affolder et al.[CDF Coll.], Phys. Rev. Lett.  {85} (2000)
2886.
\bibitem{33}F.~Yuan, K-T.~Chao, Phys. Rev.  {D63} (2001) 034006.
\bibitem{34}F.~Yuan, K-T. Chao, Phys. Rev. Lett.  {D87} (2001)
022002-L.
\bibitem{35}Ph.~H$\ddot a$gler et al., Phys. Rev. Lett.  {86} (2001)
1446.
\bibitem{36}Ph.~H$\ddot a$gler et al., Phys. Rev.  {D63} (2001)
077501.
\end{thebibliography}
\end{document}